\documentclass[aps,prl,unsortedaddress,superscriptaddress,longbibliography,reprint]{revtex4-2}
\usepackage{hyperref}
\usepackage{graphicx}
\usepackage{amsmath}
\usepackage{amssymb}
\usepackage{times,mathptmx}
\usepackage{dcolumn}
\usepackage{ifthen}
\usepackage{color}
\usepackage{proof}

\begin{document}

\title[]{Electrically induced angular momentum flow between separated ferromagnets}


\author{Richard Schlitz}
    \email{richard.schlitz@mat.ethz.ch}
    \affiliation{Department of Materials, ETH Z\"urich, 8093 Z\"urich, Switzerland}
\author{Matthias Grammer}
    \affiliation{Walther-Mei{\ss}ner-Institut, Bayerische Akademie der Wissenschaften, 85748 Garching, Germany}
    \affiliation{Physics Department, TUM School of Natural Sciences, Technische Universit\"at M\"unchen, 85747 Garching, Germany}
\author{Tobias Wimmer}
    \affiliation{Walther-Mei{\ss}ner-Institut, Bayerische Akademie der Wissenschaften, 85748 Garching, Germany}
    \affiliation{Physics Department, TUM School of Natural Sciences, Technische Universit\"at M\"unchen, 85747 Garching, Germany}
\author{Janine G\"{u}ckelhorn}
    \affiliation{Walther-Mei{\ss}ner-Institut, Bayerische Akademie der Wissenschaften, 85748 Garching, Germany}
    \affiliation{Physics Department, TUM School of Natural Sciences, Technische Universit\"at M\"unchen, 85747 Garching, Germany}
\author{Luis Flacke}
    \affiliation{Walther-Mei{\ss}ner-Institut, Bayerische Akademie der Wissenschaften, 85748 Garching, Germany}
    \affiliation{Physics Department, TUM School of Natural Sciences, Technische Universit\"at M\"unchen, 85747 Garching, Germany}
\author{Sebastian T.B. Goennenwein}
    \affiliation{Department of Physics, University of Konstanz, 78457 Konstanz, Germany}
\author{Rudolf Gross}
    \affiliation{Walther-Mei{\ss}ner-Institut, Bayerische Akademie der Wissenschaften, 85748 Garching, Germany}
    \affiliation{Physics Department, TUM School of Natural Sciences, Technische Universit\"at M\"unchen, 85747 Garching, Germany}
    \affiliation{Munich Center for Quantum Science and Technology (MCQST), 80799 M\"unchen, Germany}
\author{Hans Huebl}
    \affiliation{Walther-Mei{\ss}ner-Institut, Bayerische Akademie der Wissenschaften, 85748 Garching, Germany}
    \affiliation{Physics Department, TUM School of Natural Sciences, Technische Universit\"at M\"unchen, 85747 Garching, Germany}
    \affiliation{Munich Center for Quantum Science and Technology (MCQST), 80799 M\"unchen, Germany}
\author{Akashdeep Kamra}
	\affiliation{Condensed Matter Physics Center (IFIMAC) and Departamento de F\'{i}sica Te\'{o}rica de la Materia Condensada, Universidad Aut\'{o}noma de Madrid, E-28049 Madrid, Spain}
\author{Matthias Althammer}
    \email{matthias.althammer@wmi.badw.de}
    \affiliation{Walther-Mei{\ss}ner-Institut, Bayerische Akademie der Wissenschaften, 85748 Garching, Germany}
    \affiliation{Physics Department, TUM School of Natural Sciences, Technische Universit\"at M\"unchen, 85747 Garching, Germany}

\date{\today}

\begin{abstract} 
Converting angular momentum between different degrees of freedom within a magnetic material results from a dynamic interplay between electrons, magnons and phonons. 
This interplay is pivotal to implementing spintronic device concepts that rely on spin angular momentum transport.
We establish a new concept for long-range angular momentum transport that further allows to address and isolate the magnonic contribution to angular momentum transport in a nanostructured metallic ferromagnet.
To this end, we electrically excite and detect spin transport between two parallel and electrically insulated ferromagnetic metal strips on top of a diamagnetic substrate.
Charge-to-spin current conversion within the ferromagnetic strip generates electronic spin angular momentum that is transferred to magnons via electron-magnon coupling.
We observe a finite angular momentum flow to the second ferromagnetic strip across a diamagnetic substrate over micron distances, which is electrically detected in the second strip by the inverse charge-to-spin current conversion process.
We discuss phononic and dipolar interactions as the likely cause to transfer angular momentum between the two strips. 
Moreover, our work provides the experimental basis to separate the electronic and magnonic spin transport and thereby paves the way towards magnonic device concepts that do not rely on magnetic insulators.
\end{abstract}
\maketitle

The advent of spintronics is usually associated with spin transport mediated by charge carriers in metallic ferromagnets (FMs) and related effects such as the giant ~\cite{Binasch_GMR,Baibich_GMR,Parkin_GMR} and the tunneling magnetoresistance~\cite{Julliere_TMR,Miyazaki_TMR,Moodera_TMR}. These phenomena are based on the spin of itinerant electrons and the flow of spin-polarized electric currents in FMs. A substantial step forward was taken with the transfer of spin angular momentum from itinerant electrons to the magnetic order via spin transfer torque (STT) effects~\cite{slonczewski_current-driven_1996,berger_emission_1996,ralph_spin_2008}. This has enabled more efficient magnetic memories and the generation of quantized excitations of the magnetic system (magnons) at GHz frequencies in spin transfer torque oscillators by a DC charge current bias~\cite{Tsoi_PRL_1998,Myers1999,Katine_2000,Kiselev2003}, which are now considered as spintronic building blocks for computing architectures beyond the von Neumann scheme~\cite{csaba_computational_2013,locatelli_spin-torque_2014,grollier_spintronic_2016,romera_vowel_2018}. 

A key breakthrough came with the demonstration that spin information can be communicated via magnons instead of itinerant electrons. Magnons are bosonic quasiparticles and thus provide intriguing possibilities such as condensation~\cite{Demokritov2006,Pirro2021} and quantum fluctuations engineering~\cite{Kamra2020,Pirro2021,Yuan2022}, not admitted by the fermionic electrons. The recently developed pathway for spin transport via magnons is to use magnetic insulators, where itinerant electron transport is absent. Nevertheless, to enable electrical access, non-magnetic metal strips on top of magnetic insulators have been used to electrically induce magnon spin transport~\cite{CornelissenMMR,SchlitzMMR,KleinMMR,Klaui2018}. In these heterostructures, one utilizes charge-to-spin current interconversion in the non-magnetic metal, e.g., by the direct and inverse spin Hall effect (SHE)~\cite{Dyakonov1971,Hirsch1999,Hoffmann2013,Sinova2015}, to inject and detect magnon spin transport in the adjacent magnetic insulator. This approach is based on the transfer of angular momentum across the interface which therefore needs to be transparent to spin currents.

\begin{figure*}[t]
\includegraphics[width=170mm]{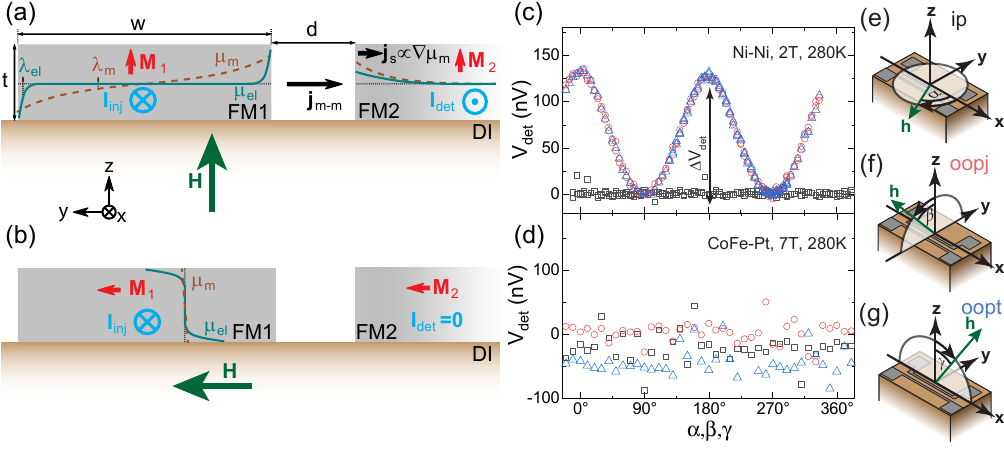}
    \caption{
    (a) Spin transport experiments between two FM strips on a diamagnetic insulator (DI). Charge-to-spin conversion leads to an electronic spin accumulation $\mu_\mathrm{el}$ at the edges of the FM1 layer upon application of a current $I_\mathrm{inj}$ with the spin polarization oriented along the magnetization $\mathbf{M}$. The electronic spin accumulation gives rise to a magnon chemical potential $\mu_\mathrm{m}$, which relaxes on the scale of the magnon diffusion length $\lambda_\mathrm{m}$, which exceeds the electronic spin diffusion length $\lambda_\mathrm{el}$. The finite magnon accumulation at the right side couples to the thermal magnon bath in the FM2 strip by dipolar coupling and via phonon spin transport through the DI and thereby induces a dc current $I_\mathrm{det}$ in the FM2 strip by spin-to-charge conversion. (b) Illustration of the chemical potential profiles for $\mathbf{M}\parallel \mathbf{y}$ . The electron spin accumulation is now at the top and bottom surfaces and fails to generate any substantial magnon chemical potential as $t\ll\lambda_\mathrm{m}$. Thus, no angular momentum transport signal is observed in the detector.
    (c,d) Angle-dependence of $V_\mathrm{det}$ for Ni-Ni strips (panel c) 
and CoFe-Pt strips (panel d) 
obtained at $280\;\mathrm{K}$ and $\mu_0 H=2\;\mathrm{T},7\;\mathrm{T}$. Black squares correspond to ip-rotations, red circles to oopj-rotations and blue triangles to oopt rotations, respectively. (e,f,g) Illustration of the rotation planes used for the angle-dependent measurements:  in-plane rotation (e, ip), out-of-plane perpendicular to $\mathbf{j}$ (f, oopj), and out-of-plane perpendicular to $\mathbf{t}$ (g, oopt).}
\label{FigTheory}
\end{figure*}

In this article, we develop and demonstrate a pathway to access magnonic spin transport in an all electrical fashion without requiring magnetic insulators. To realize this concept, we utilize two isolated ferromagnetic (FM) metal strips with width $w$ and separation $d$ as electrical injector and detector of spin current while allowing angular momentum transport between the two strips [see Fig.~\ref{FigTheory}(a,b)]. The two FM strips are deposited on top of a diamagnetic insulator (DI) as detailed in the Supplemental Material (SM)~\footnote{\label{fn:supp}See Supplemental Material at [url], which includes Refs.~\cite{SchreierCSSE,Chuang2017,Gao_2022,Ky1966,Ky1967,Pu2006,Avery2012,Omori2019_AHE,Wimmer2019_SAHE}, for details on the fabrication of the structures, electrical connection scheme, data analysis and fit results. Moreover, we discuss there the thermopower signals, additional measurements on other ferromagnetic strips and measurements in a 4 strip geometry.}.
Fig.~\ref{FigTheory}(c) shows that a Ni-Ni structure exhibits a finite voltage $V_\mathrm{det}$ on one Ni wire, when driving a current $I_\mathrm{inj}$ in the adjacent Ni strip and applying a sufficiently large magnetic field $H$ to align the magnetization $M$ along the surface normal $\mathbf{z}$.
Rotating the magnetic field in three orthogonal rotation planes with the rotation angles $\alpha, \beta, \gamma$ [see Fig.~\ref{FigTheory}(e)-(g) for the definition of the rotation planes], we observe that $V_\mathrm{det}$ vanishes to within our experimental resolution when the magnetization is oriented in the sample plane. Note that we utilize a current reversal scheme in our measurements to rule out any contributions from thermovoltages arising due to Joule heating from $I_\mathrm{inj}$(see SM~\footnotemark[1] for a detailed description).

To elucidate the origin and shape of $V_\mathrm{det}$, we first consider the situation in the injector wire (FM1) with the magnetization $\mathbf{M}_1$ oriented along the surface normal as illustrated in Fig.~\ref{FigTheory}(a).
By charge-to-spin current conversion via the anomalous spin Hall effect (SHE)~\cite{Sinova2015,Taniguchi2015_SAHE,Das2017,Das2018}, the charge current bias $I_\mathrm{inj} \parallel \mathbf{x}$ in the FM1 strip generates an electronic spin accumulation polarized along the magnetization direction at the two sides of the wire over 
the electronic spin decay length $\lambda_\mathrm{el}$. This can be seen as the generation of electronic spin accumulation via the typical SHE~\cite{Hirsch1999,Sinova2015}, like in a non-magnetic metal, but with a crucial difference: the inherent magnetization eliminates all spin accumulation transverse to it~\cite{Taniguchi2015_SAHE,Das2017,Das2018}.
The electronic spin accumulation (described via the electron spin chemical potential $\mu_\mathrm{el}$) interacts with the thermal magnon bath via inelastic spin-flip scattering. 
A magnon accumulation is thereby generated on the edges over the magnon decay length $\lambda_\mathrm{m}$ [see Fig.~\ref{FigTheory}(a)]~\cite{Cheng2017}.

The (spin) angular momentum of the magnon accumulation at the right side of FM1 can be transported to the detector wire (FM2) via dipolar coupling~\cite{Dzyapko_2013,Sadovnikov2015} and phonon-mediated spin transport through the DI~\cite{Streib2018,holanda_detecting_2018,holanda_evidence_2021,An2020,Rckriegel2020,Rckriegel2020b,rezende2021}.
A nonzero angular momentum transfer between the two separated FMs on the DI can therefore take place so that an angular momentum current $\mathbf{j}_\mathbf{m-m}$ flows along $-\mathbf{y}$ into the FM2 strip.
Therefore, a finite gradient in the magnon chemical potential at the left side of FM2 arises, which in turn leads to an electronic spin current flow $\mathbf{j}_\mathbf{s} = - j_\mathbf{s} \mathbf{y}$.
In the FM2 strip $\mathbf{j}_\mathbf{s}$ is then transformed into a charge current $I_\mathrm{det}$ flowing along $-\mathbf{x}$ by the inverse charge-to-spin current conversion processes. 
Since in the experiment we detect the open circuit voltage $V_\mathrm{det}$, the resulting charge accumulation leads to an electric field $\mathbf{E}^\mathrm{det}\parallel \mathbf{x}$.

This picture is consistent with the experimentally observed electric field direction.
We emphasize that the direction of the observed electric field is opposite compared to the typical lateral magnon spin transport in magnetic insulators~\cite{CornelissenMMR,SchlitzMMR,Klaui2018,KleinMMR}. 

The magnetic field orientation dependence of the signal can be motivated when considering the device geometry [see Fig.~\ref{FigTheory}(b)]. 
The distance between and the width of the wires is always large compared to their thickness, i.e. $t\ll d,w$.
If we change the orientation of the magnetization into the sample plane $\mathbf{M}_1 \parallel \mathbf{y}$, the charge-to-spin current conversion process via the anomalous SHE gives rise only to a spin accumulation at the top and bottom surfaces of the metal wires on the length scale of $\lambda_\mathrm{el}$.
The associated spatial variation of the magnon chemical potential is vanishingly small, as a significant variation is possible only on a length scale $\lambda_m \gg t$. 
As the magnon chemical potential at opposite film surfaces has opposite sign, to a good approximation the generated magnon chemical potential vanishes and no transfer of angular momentum to FM2 is possible. 
As nonequilibrium magnons are generated in FM1 only when the magnetization has a finite component along the surface normal, our proposed mechanism is consistent with a $\cos^2(\beta,\gamma)$ modulation of the signal taking into account the projection of the magnetization on the surface normal for both the injector and detector wire, where the same process takes place in reverse. This is consistent with the experimental result shown in Fig.~\ref{FigTheory}(c). We note that a similar transport of out-of-plane spins was observed in all-electrical magnon transport experiments where ferromagnetic wires are placed upon magnetic insulators~\cite{Das2018}. However, the absence of magnons in the diamagnetic insulator excludes such a mechanism in our devices.

From the angular dependence we extract the modulation amplitude $\Delta V_\mathrm{det}$ as shown in Fig.~\ref{FigTheory}(c). For the Ni-Ni strips $\Delta V_\mathrm{det}$ is of the order of $100\;\mathrm{nV}$, corresponding to an equivalent resistance $\approx1\;\mathrm{m\Omega}$ for this device. 
This signal magnitude is comparable to the amplitudes in all-electrical magnon transport experiments using FM strips and yttrium iron garnet layers~\cite{Wimmer2019_SAHE,Das2017,Das2018}, showing that such signals can be routinely detected. 
The absence of the effect when one of the electrodes is replaced by Pt shows that a purely electronic spin accumulation is insufficient to observe the signal as shown in Fig.~\ref{FigTheory}(d). 
This observation indicates that the effect relies on the interconversion of electronic and magnonic spin accumulations.
Furthermore, a contribution of potential orbital currents can be excluded by the same notion, as a finite orbital Hall effect is attributed to Pt.

\begin{figure}[t]
\includegraphics[width=85mm]{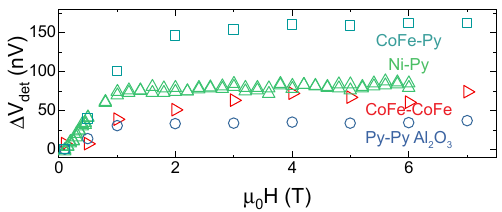}
\caption{
External magnetic field dependence of $\Delta V_\mathrm{det}$ for Py-Py($t=15\,\mathrm{nm}$) strips 
 (red symbols), for CoFe-CoFe strips 
 (light blue symbols), and Ni-Py strips
 (green symbols), and CoFe-Py strips 
 (brown symbols) at $T=280\;\mathrm{K}$. For large external magnetic fields $\Delta V_\mathrm{det}$ saturates once the external magnetic field is large enough to fully align the magnetization of both strips along the out-of-plane direction.
 }
\label{Fig_FieldDep}
\end{figure}

The magnetic field dependence of $\Delta V_\mathrm{det}$ corroborates the importance of the out-of-plane projection of the magnetization (see Fig.~\ref{Fig_FieldDep}), as it exhibits a saturation behaviour for all FM1-FM2 strips within the noise level of our measurements. 
The onset of saturation agrees with the external magnetic field required to overcome the magnetic anisotropy and thus align the magnetization along $\mathbf{z}$. 
Moreover, since $\Delta V_\mathrm{det}$ is not reduced for large magnetic fields, we conclude that thermal magnons with large wavevector dominantly contribute to the spin transport. This is further supported by measurements with different ferromagnetic materials for FM1 and FM2, where the fundamental magnon mode frequency differs.

\begin{figure}[t]
\includegraphics[width=85mm]{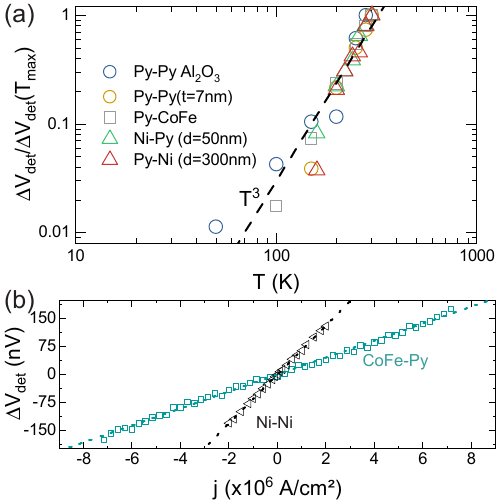}
\caption{
(a) Extracted temperature dependence for several FM1-FM2 devices on different samples. 
 The extracted values exhibit a $T^{3}$-dependence. (b) Extracted current bias dependence (using the charge current density $j$ calculated from $I_\mathrm{inj}$) of $\Delta V_\mathrm{det}$ for Ni-Ni strips (
$280\;\mathrm{K}$, $2\;\mathrm{T}$, brown symbols), 
 and for CoFe-Py strips (
$280\;\mathrm{K}$, $7\;\mathrm{T}$, dark blue symbols). A linear fit with zero intercept is indicated for each data by a dotted line.}
\label{Fig_eval}
\end{figure}

To establish the dominant angular momentum transport mechanism, we investigate the temperature, drive current and injector-detector distance dependence of $\Delta V_\mathrm{det}$. 
Transfer of angular momentum from electrons to magnons depends on the thermal magnon population in the ferromagnet and thus on its temperature~\cite{Zhang_Zhang_PRB_2012,CornelissenTheory}. 
This can be intuitively understood as the coupling strength between electrons and magnons scales with the magnon mode occupancy, which increases with temperature.
As shown in Fig.~\ref{Fig_eval}(a), $\Delta V_\mathrm{det}$ (amplitude of the $\cos^2$-dependence) indeed depends sensitively on temperature.
$\Delta V_\mathrm{det}$ decreases with decreasing temperatures for all investigated samples. 
For $T \leq 100\;\mathrm{K}$, the extracted $\Delta V_\mathrm{det}$ is below our detection limit. 
This observation is consistent with the magnonic origin and the expected role of the thermal magnon bath in the interconverison between the electronic and magnonic spin.
Note that for better comparison between the different combinations of the FMs we normalized $\Delta V_\mathrm{det}(T)$ to the value $\Delta V_\mathrm{det}$ at the maximum temperature in the measurements ($280\;\mathrm{K}$ or $300\;\mathrm{K}$). 

The temperature dependence is similar to all-electrical magnon transport experiments, where a reduction at low temperatures is found~\cite{SchlitzMMR,CornelissenTemp}. 
However, the $T^{\alpha}$ dependence with $\alpha\sim1.5$ observed in Ref.~\cite{SchlitzMMR} and which is attributed to the spin convertance at the interface~\cite{Zhang_Zhang_PRB_2012,Bender2015} to the spin transport does not fit our data.
The dashed black line indicates that $\alpha\sim3$ is more adequate to describe our data.
The implications of this dependence on the dominant mechanism for angular momentum transport are discussed further below.

As shown in Fig.~\ref{Fig_eval}(b), $\Delta V_\mathrm{det}$ depends dominantly linear on $I_\mathrm{inj}$ for all FM1-FM2 strips (dashed same colored lines indicate a linear fit with zero $y$-axis intercept). 
This is compatible with a non-equilibrium process in linear response driven by electronic and magnonic spin accumulations proportional to the electrical current drive~\cite{CornelissenTheory}. 
For large currents a cubic deviation is observed for CoFe-CoFe that stems from Joule heating increasing the device temperature and thereby $\Delta V_\mathrm{det}$.
Note that thermal effects to the voltage that are independent of the direction of current flow are removed by a current reversal technique and therefore do not contribute to the signal (see Methods).

\begin{figure}[t]
\includegraphics[width=85mm]{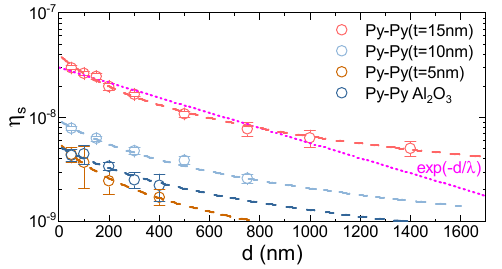}
\caption{
The spin transfer efficiency $\eta_\mathrm{s}$ as a function of the distance $d$ is shown for 4 different Py samples with varying DI substrate and Py thickness ($2.5\;\mathrm{T}$, $280\;\mathrm{K}$). The dashed lines represent a fit to each data set using Eq.~\eqref{eq:AkashDistance}, which are in excellent agreement with the obtained data. The dotted magenta line shows an exponential decay fit to the Py-Py(15) data for comparison, yielding a decay length $\lambda \approx 600\,\mathrm{nm}$.
 }
\label{Fig4_distance}
\end{figure}

To analyze how the wire separation $d$ of the FM1-FM2 strips and the thickness $t$ influence $\Delta V_\mathrm{det}$, we utilized 4 samples with several Py-Py strips with identical width $w$ and a systematic variation in $d$, while varying $t$ for each sample deposited on YAG or Al$_2$O$_3$ substrates.
To remove any device variations, we calculated the spin transfer efficiency $\eta_\mathrm{s}(d)=I_\mathrm{det}/I_\mathrm{inj}=\Delta V_\mathrm{det}/V_\mathrm{inj}$ for each device. Here, we utilized that the resistance of our injector and detector strips are identical within $1-3$ \%. 
For $\eta_\mathrm{s}(d)$ we observe a reduction with $d$ (see Fig.~\ref{Fig4_distance}) for all investigated samples. Moreover, our data suggest that the spin transfer efficiency is enhanced with increasing $t$, but the investigated thickness range at present is too limited to extract a concrete dependence. 

The $d$ dependence of the signal may provide valuable information about the physical mechanisms underlying our observed finite magnon conductance across the spatially separated FM1-FM2. The contribution due to the diffusive transport of an intermediary, such as phonons, may be expected to result in an exponential dependence $\sim \exp(-d/\lambda_{\mathrm{ph}})$\footnote{In principle, diffusive transport at length scales well below the decay length can manifest a 1/d dependence~\cite{CornelissenMMR}. However, this is feasible only when the injector and detector strongly couple to the intermediary particle. Such a strong coupling between the two magnonic systems via phonons seems highly unlikely in our devices.}. We note that an oscillatory behavior has been predicted~\cite{Rckriegel2020} for the angular momentum transport mediated by phonons between two ferromagnets and for short distances $d$, which we do not observe in our experiments.
On the other hand, magnons in either ferromagnet can extend to the respective other one by dipolar coupling, so that angular momentum transfer between the two FMs may manifest with an algebraic decay with $d$. This phenomenon is established in literature and is referred to as magnon tunneling~\cite{schneider_spin-wave_2010,sadovnikov_frequency_2016,wang_magnonic_2020}. Quantitative modeling of this mechanism without knowledge of the spectral distribution of contributing magnons is however complicated and rendered analytically intractable by the finite sizes of the two FMs and the long-range nature of the dipolar interaction. Thus, we heuristically motivate a $d$ dependence that may capture the dipolar coupling contribution. 
	
The dipolar interaction between two thin magnetic wires separated by distance $d$ scales as $\sim 1/d^2$. The corresponding coupling between an infinitely wide film and a wire placed at distance $d$ from its edge scales as $\sim 1/d$. Our finite size FMs may be expected to manifest a scaling $\sim 1/d^n$, with $1 \leq n \leq 2$. Furthermore, considering that our FMs are wide and the relevant angular momentum transfer originates predominantly from a region about $\lambda_m$ from the FM edge, the effective separation between the wires becomes $d + c_2$ with $c_2$ a distance comparable to $\lambda_m$. 
Thus, we may consider the following distance dependence for the dipolar mechanism, assuming $n = 1$ for simplicity~\footnote{Here, in principle, we could have used $n$ as an additional fitting parameter. However, a significantly larger data set than what we have obtained in this study is needed to reliably include $n$ as a fitting parameter.}:
\begin{equation}
    \eta_\mathrm{s}(d)=\frac{c_1}{d+c_2},
    \label{eq:AkashDistance}
\end{equation}
where $c_1$ is a scaling constant. We fit our data with  Eq.\eqref{eq:AkashDistance} (dashed lines in Fig.~\ref{Fig4_distance}) and obtain similar values $c_2\approx200\;\mathrm{nm}$ for all investigated samples (see SM~\footnotemark[1] for a detailed list of all fit parameters), suggesting that the dipolar coupling is likely the main contribution to our observed angular momentum transfer. For comparison, we also show the exponential fit to our data in Fig.~\ref{Fig4_distance} for the Py-Py($t=15\;\mathrm{nm}$) dataset. Furthermore, in the dipolar interaction-mediated transport scenario, our observed temperature scaling with $T^{3}$ (Fig.~\ref{Fig_eval}a) is attributed to the product of the magnon number in each FM, which both follow the Bloch $T^{3/2}$ dependence.

In summary, our results establish a device architecture to investigate and exploit magnonic angular momentum transport between two FM strips by all-electrical means. The concept is based on charge-to-spin current conversion in the metallic ferromagnets and the interplay of the accumulated electron spin with the thermal magnon gas. 
The observed scaling with distance and temperature is compatible with angular momentum transport by dipolar coupling between the two ferromagnets, but we cannot rule out the potential dominance or relevance of a phononic contribution. Our results lay the foundation for all-electrical investigation of magnon spin transport and charge-to-spin current conversion processes in nano-structured FMs. The presented scheme provides a blueprint for magnonic devices that do not rely on magnetic insulators and transparent interfaces for spin currents to separate the electronic and magnonic systems. Our results further provide a perspective on the characterization and exploitation of magnonic crystals~\cite{chumak_magnonic_2017} for spin wave propagation.

\section{Acknowledgements}
M.G., T.W., J.G., L.F., H.H., R.G., and M.A. acknowledge financial support by the German Research Foundation (Deutsche Forschungsgemeinschaft, DFG) under Germany’s Excellence
Strategy -- EXC-2111 -- 390814868. S.T.B.G. acknowledges financial support by the German Research Foundation (Deutsche Forschungsgemeinschaft, DFG) via the SFB 1432 – Project-ID 425217212. A.K. acknowledges financial support from the Spanish Ministry for Science and Innovation -- AEI Grant CEX2018-000805-M (through the ``Maria de Maeztu'' Programme for Units of Excellence in R\&D) and grant RYC2021-031063-I funded by MCIN/AEI/10.13039/501100011033 and ``European Union Next Generation EU/PRTR''. We thank G.E.W.~Bauer for fruitful discussions during the development of this manuscript and T.~Helm for experimental support.

\bibliography{Bibliography}
\end{document}